\documentclass[11pt,twoside]{article}
\usepackage{asp2010}

\resetcounters

\begin{document}

\title{Herschel observations: contraints on dust attenuation and star formation histories at high redshift}
\author{Denis~Burgarella,$^1$V\'eronique~Buat,$^1$ Carlotta~Gruppioni,$^2$, Olga~Cucciati,$^2$, S\'ebastien~Heinis,$^1$ and the PEP/HerMES Team
\affil{$^1$Aix-Marseille Universit\'e, CNRS, LAM (Laboratoire d'Astrophysique de Marseille) UMR 7326, 13388, Marseille, France}
\affil{$^2$INAF-Osservatorio Astronomico di Bologna, via Ranzani 1, 40127, Bologna, Italy}}

\begin{abstract}
 {\sl SPICA} is one of the key projects for the future. Not only its instrument suite will open up a discovery window but they will also allow to physically understand some of the phenomena that we still do not understand in the high-redshift universe. Using new homogeneous luminosity functions (LFs) in the Far-Ultraviolet (FUV) from VVDS and in the Far-Infrared (FIR) from {\sl Herschel}/PEP and {\sl Herschel}/HerMES, we studied the evolution of the dust attenuation with redshift. With this information, we are able to estimate the redshift evolution of the total (FUV + FIR) star formation rate density (${\rm SFRD_{TOT}}$). Our main conclusions are that: 1) the dust attenuation $A_{FUV}$ is found to increase from $z = 0$ to $z \sim 1.2$ and then starts to decrease until our last data point at $z = 3.6$; 2) the estimated SFRD confirms published results to $z \sim 2$. At $z > 2$, we observe either a plateau or a small increase up to $z \sim 3$ and then a likely decrease up to $z = 3.6$; 3) the peak of ${\rm A_{FUV}}$ is delayed with respect to the plateau of ${\rm SFRD_{TOT}}$ but the origin of this delay is not understood yet, and {\sl SPICA} instruments will provide clues to move further in the physical understanding of this delay but also on the detection and redshift measurements of galaxies at higher redshifts. This work is further detailed in \citep[][]{Burgarella_2013}.
\end{abstract}

\section{Introduction}

Over the past 15 years or so, astronomers have tried to measure the evolution of the cosmic star formation rate density (SFRD) moving higher and higher in redshift. However, we quickly understood that one of the main issues was to account for the total SFRD and not only for the far-ultraviolet (FUV) one. This means either a dust correction of the FUV SFRD or, better, a measure of the total i.e., FUV plus far-infrared (FIR = bolometric IR) SFRD. Knowing how the dust attenuation evolves in redshift is therefore mandatory if one wishes to study the redshift evolution of the SFRD.

\cite{Takeuchi_2005} estimated the cosmic evolution of the SFRD from the FUV and FIR. An increase of the fraction of hidden SFR is found to $z = 1$ where it reaches ~84\%. The dust attenuation increases from A$_{FUV} \sim 1.3$ mag locally to A$_{FUV} \sim 2.3$ mag at $z = 1$. From the FUV only \cite{Cucciati_2012} show that the mean dust attenuation A$_{FUV}$ agrees with \citet{Takeuchi_2005} over the range $0 < z < 1$, remains at the same level to $z \sim 2$, and declines to $\sim$ 1 mag at $z \sim 4$. 

Using FUV luminosity functions (LFs) published in \cite{Cucciati_2012} and FIR LFs from {\sl Herschel}\footnote{From two {\sl Herschel} Large Programmes: PACS Evolutionary Probe \citep[PEP,][]{Lutz_2011} and the {\sl Herschel} Multi-tiered Extragalactic Survey \citep[HerMES,][]{Oliver_2012}} \citep{Gruppioni_2013}, we are able to constrain the redshift evolution of  $\log_{10} (L_{FIR} / L_{FUV}$) (aka $IRX$) to $z \sim 4$ for the first time directly from FIR data. With this information, we can estimate the redshift evolution of $\rho_{FIR}/\rho_{FUV}$ as well as $\rho_{TOT} = \rho_{FIR} + \rho_{FUV}$. 

The information gathered in this work poses a number of questions that would be addressed by  {\sl SPICA}. The physical study with  {\sl SPICA} of the high redshift galaxies detected by Herschel is crucial to better understand the formation and evolution of galaxies.

Throughout this paper we adopt a $\Lambda$CDM cosmology with ($H_0, \Omega_m, \Omega_\Lambda$) = (70, 0.3, 0.7), where $H_0$ is in kms$^{-1}$Mpc$^{-1}$. All SFR and stellar masses presented assume, or have been converted to, a Salpeter IMF.

\section{About Luminosity Functions}

\begin{figure}[!ht]
\begin{center}
   \resizebox{1.0\hsize}{!}{
     \includegraphics*{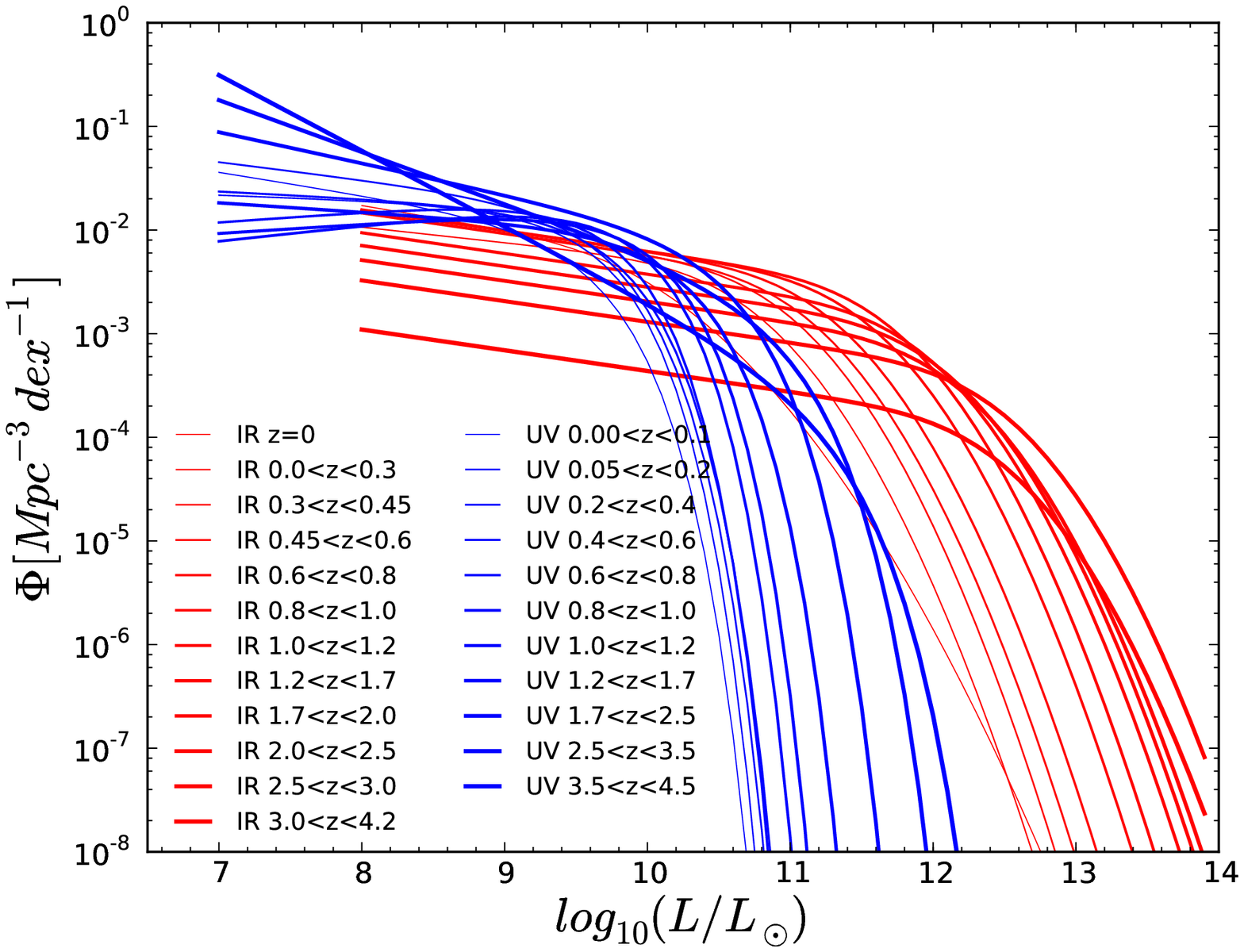}
   }
\end{center}
\caption{
Redshift evolution of the FIR \citep[red,][]{Gruppioni_2013} and FUV \citep[blue,][]{Cucciati_2012} LFs. Note that the FUV LFs are uncorrected for dust attenuation. The LFs at every other redshift are plotted in bold. The others are fainter to facilitate reading the figure. The LFs are plotted within the limits of integration.
}\label{FigLFs}
\end{figure}

Fig.~\ref{FigLFs} show the redshift variation of the LFs in FIR. The known difference in the FIR and FUV LFs  \citep[e.g.][]{Takeuchi_2005} are clearly illustrated here: bright FIR galaxies are more numerous than bright FUV galaxies at $log_{10} (L [L_\odot]) > 10$. In the FUV, except in the highest redshift bins, L$^\star$ and ${\Phi^\star}$ remain approximately constant while the faint-end slope evolves. The FIR faint end slope is not observationally constrained at high z, and \citet[][]{Gruppioni_2013} fixed it to $\alpha = 1.2$. However, L$^\star$ and ${\Phi^\star}$ were allowed to change with redshift. These different evolutions of the FUV and FIR LFs are reflected in Fig.~\ref{FigLFs} and explain the evolution of the cosmic SFRD and dust attenuation.

\section{About the Evolution of the Cosmic Dustiness}

\begin{figure}[!ht]
\begin{center}
   \resizebox{1.0\hsize}{!}{
      \includegraphics*{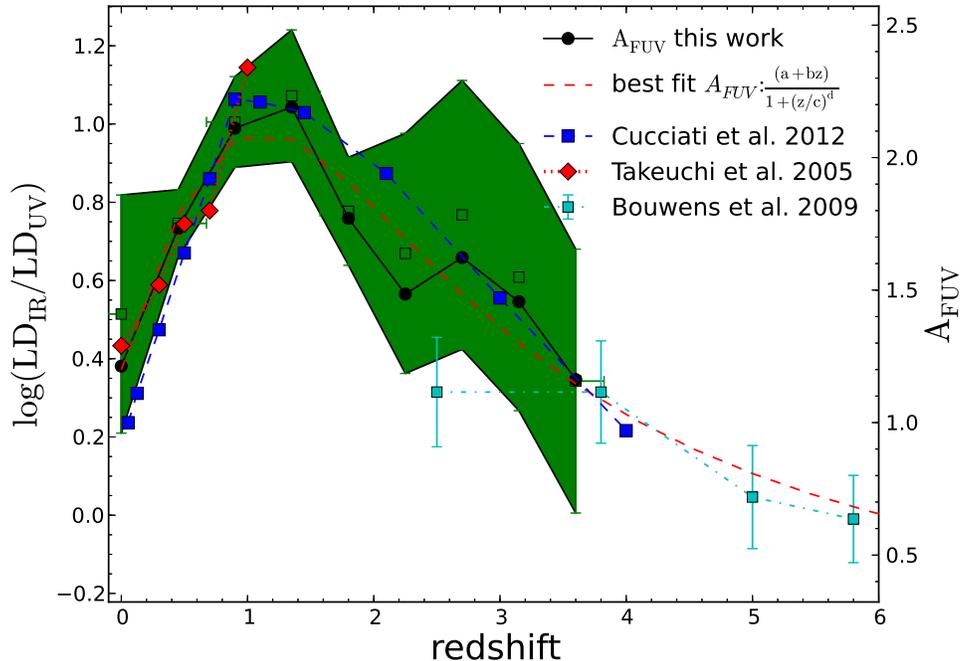}
   }
\end{center}
\caption{
Left axis: ratio of the FIR-to-FUV LDs ($IRX$). Right axis: FUV dust attenuation (A$_{FUV}$). The red dotted line with red diamonds is taken from \citep[e.g.][]{Takeuchi_2005}. The green filled area and green dots are the associated uncertainties estimated through bootstrapping with 2 000 drawings. Black dots denote the values directly computed from the LFs. At $z = 3.6$, A$_{FUV}$ reaches about the same value as at $z = 0$. \citet[][]{Takeuchi_2005} (red diamonds) used an approach identical to ours while a SED analysis (no FIR data) is performed in \citet[][]{Cucciati_2012} (blue boxes). \citet[][]{Bouwens_2009} are estimates based on the UV slope $\beta$. The limiting FUV luminosity is $10^{7} L_\odot$ or $1.65\times10^{-4} L^\star_{z=3}$. The best fit is given by ${\rm A_{FUV} (z) =\frac{(a+bz)}{1+(z/c)^d}}$ with $a = 1.20$, $b = 1.50$, $c = 1.77$, and $d = 2.19$.
}\label{FigAfuv}
\end{figure}

Fig.~\ref{FigAfuv} presents the dust attenuation in the FUV vs. z and the ratio of the FIR-to-FUV LDs integrated in the range $log_{10} (L [L_\odot$]) = [7, 14] in the FUV (i.e. L$^{min}_{FUV}$ =  $1.65\times10^{-4} L^\star_{z=3}$, \citealt[][]{Bouwens_2009}) and [8, 14] in the FIR. The FUV dust attenuation is estimated from the $IRX$ and converted to A$_{FUV}$ using \citet[][]{Burgarella_2005}\footnote{The conversion from $IRX$ to A$_{FUV}$ from \citet[][]{Burgarella_2005} is valid at $\log_{10}$ (L$_{FIR}$/L$_{FUV}$)$>$-1.2: A$_{FUV}$ = -0.028 [$\log_{10}$ L$_{FIR}$/L$_{FUV}$]$^3$ + 0.392 [$\log_{10}$ L$_{FIR}$/L$_{FUV}$]$^2$ + 1.094 [$\log_{10}$ L$_{FIR}$/L$_{FUV}$] + 0.546}. The redshift evolution of A$_{FUV}$ agrees with \citet[][]{Cucciati_2012}. Note that \citet[][]{Cucciati_2012} estimated A$_{FUV}$ through an analysis of individual SEDs up to $\lambda_{obs} = 2.2 \mu$m (Ks-band). Fig.~\ref{FigAfuv} suggests a local minimum at $z \sim 2$ that might be caused by UV-faint galaxies \citep[see Fig.~7 in][]{Cucciati_2012} that are responsible for a peak observed in the FUV LD that is not observed in the FIR. Since the fields observed in FUV and in FIR are not the same, another origin might be found in cosmic variance. The bottom line is that the existence of this trough in A$_{FUV}$ must be explored with {\sl SPICA}. Finally, higher redshift $A_{FUV}$ from the UV slope, $\beta$, suggest a continuous decline at least to $z = 6$ \citep[][]{Bouwens_2009}. 

We conclude that the cosmic dust attenuation A$_{FUV}$ reaches an absolute maximum at $z \sim 1.2$ followed by a global decline to $z = 3.6$, where it reaches about the same level as measured at $z = 0$. Beyond $z = 4$, we do not expect any increase \citep[e.g.][]{Burgarella_2005}.

\section{About The total FUV+FIR star formation density}

\begin{figure}[!ht]
\begin{center}
   \resizebox{1.0\hsize}{!}{
     \includegraphics*{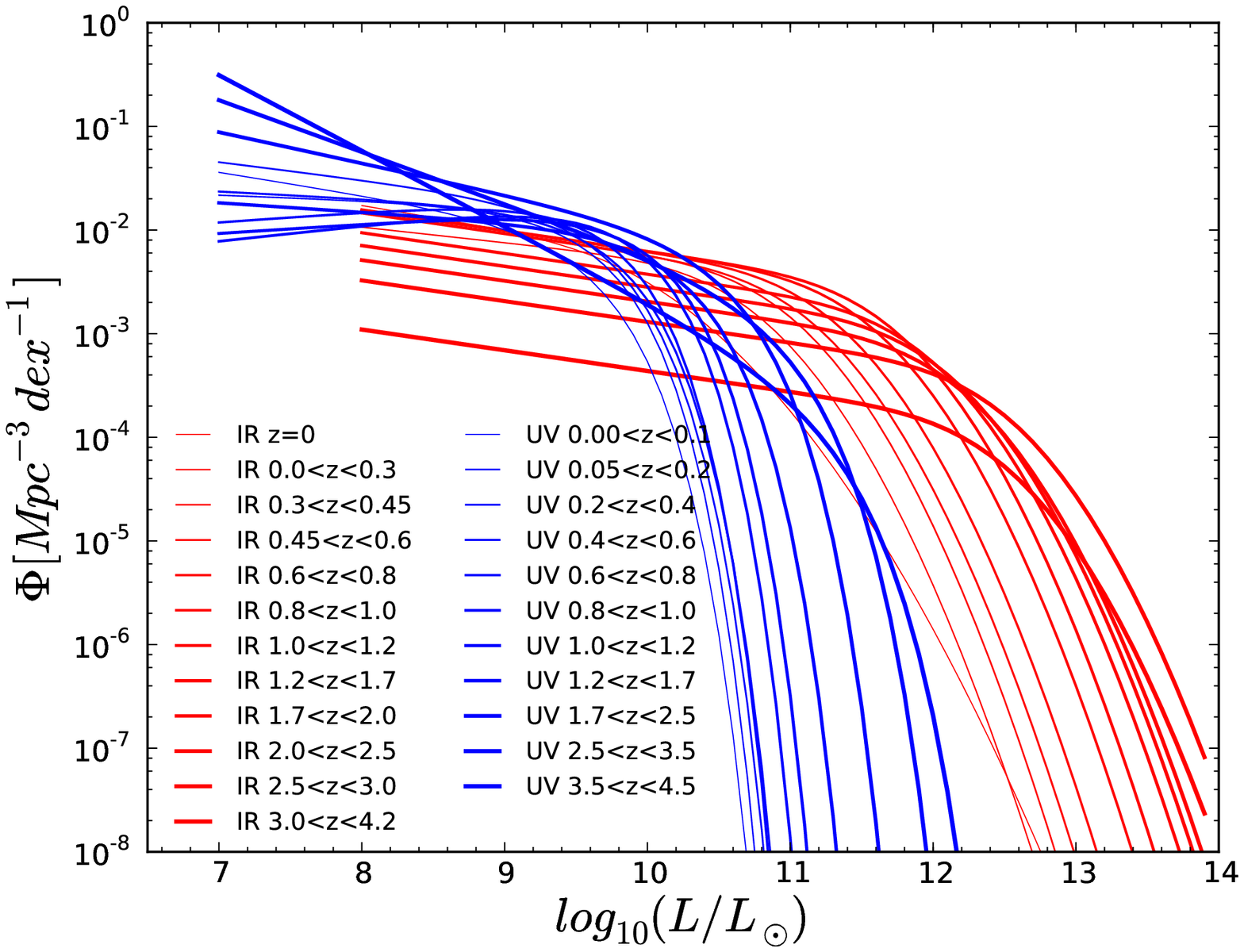}
   }
\end{center}
\caption{
SFRD densities in the FUV (blue), in the FIR (red), and in total (i.e., FUV + FIR) in green (other colors are due to overlaps of the previous colors). The lines are the mean values, while the lighter colors show the uncertainties evaluated from the 2 000 runs as in Fig.~\ref{FigAfuv}. After the initial increase of the total SFRD from $z = 0$ to $z \sim 1.2$, it remains flat or slightly increases/decreases up to $z \sim 2.5 - 3.0$ followed by a decrease. Globally and over $0 < z \le 3.6$, the total average SFRD is slightly below  that reported in \cite{Hopkins_2006} and agrees with that of \citet[][]{Behroozi_2012} up to $z \sim 2$. The SFRD from \citet[][]{Barger_2012} and preliminary results from {\sl Herschel}/SPIRE estimated by Vaccari et al. (2013, in prep.) agree with these trends. Symbols and lines are explained in the plot.
}\label{FigSFRD}
\end{figure}

Fig.~\ref{FigSFRD} suggests a flattening of the total SFRD up to $z \sim 3$ \citep[as in][]{Chary_2001, LeFloch_2005, Franceschini_2010, Goto_2010, Magnelli_2012}, where the UV data favor a peak followed by a decrease. Note that we cannot rule out a small increase or decrease within the uncertainties. All in all, our total SFRD agrees fairly well with that of \cite{Hopkins_2006} in the same redshift range. However, discrepancies exist: our total SFRD is lower at $z < 1$ and is only marginally consistent, but lower, at $z > 3$. Moreover, PACS data are less sensitive at higher than at lower redshift because the rest-frame wavelength moves into the mid-IR. The preliminary FIR SFRD from Vaccari et al. (2013, in prep.) ({\sl Herschel}/SPIRE selection) agrees excellently over the $0 < z \le 2$ range, but is slightly higher than that derived from PACS at $z > 3$. However, this is only a $\sim 2\sigma$ difference. \cite{Barger_2012} published a FIR SFRD based on SCUBA-2 data that also agrees with ours at $2 < z < 4$. We first tried to fit  ${\rm SFRD_{TOT}}$ with a one-peak analytical function \citep[e.g.][]{Hopkins_2006, Behroozi_2012}, but the results are not satisfactory. So, we combined two Gaussians, $$a_1 e^{\frac{-(z-z_1)^2}{2 \sigma_1^2}} + a_2 e^{\frac{-(z-z_2)^2}{2 \sigma_2^2}},$$ with $a_1 = 0.1261 \pm 0.0222$, $\sigma_1 = 0.5135 \pm 0.0704$, $z_1 = 1.1390 \pm 0.0959$ and $a_2 = 0.2294 \pm 0.0222$, $\sigma_2 = 0.8160 \pm 0.0964$, $z_2 = 2.7151 \pm 0.0839$. At higher redshifts, we made assumptions that are explained below.

The cosmic SFRD presents a (weak) maximum at $z \sim 2.5 - 3.0$ (i.e., between 2.6 - 2.1 Gyr) while the dust attenuation presents a maximum at $z \sim 1.2$ (i.e. 5 Gyr). We tried to lock the faint-end slope of the UV LF -1.2, to see how far out in redshift the obscuration peak could potentially move, but we detected no change, suggesting this effect is solid. We have no definite explanation for this delay of $\sim$ 2.7 Gyr. Type II supernovae start producing dust earlier than AGB stars \citep[e.g. Fig. 3 in][]{Valiante_2009} but the difference in timescales is too short and only on the order of a few 10 Myr for the onset of dust formation. Dust grain destruction in the ISM might play a role \citep[e.g.][]{Dwek_2011} but the efficiency of destruction is only poorly known and depends on the star formation history. These dust-related origins for the delayed maximum are un-likely. The best explanation might be that this delay is related to a global move of galaxies in the [$\log_{10} (L_{FIR} / L_{FUV}$) vs. $\log_{10} (L_{FIR} + L_{FUV})$] diagram. \citet[][]{Buat_2009} showed that galaxies evolve in redshift from $z = 0$ to $z = 2$ in this diagram, with high-redshift sources having lower $IRX$ at given total luminosities. 

\section{Conclusions}

On the one hand, the variation of the cosmic dust attenuation with redshift suggests a peak in the dust attenuation at $z \sim 1.2$ followed by a decline to $z = 3.6$. On the other hand, the total (FUV+FIR) cosmic SFRD increases from $z = 0$ to $z \sim 1.2$, remains flat to $z \sim 2.5 - 3.0$ followed by a decrease at higher redshifts and reaches the same level at $z \sim 5 - 6$ as is measured locally if we assume no variations in this trend. 

So, the SFRD and A$_{FUV}$ do not exactly follow the same trends as seen in Fig.~\ref{FigAfuv} and Fig.~\ref{FigSFRD}. The peak of the dust attenuation is delayed with respect to the plateau of the total SFRD by about 3 Gyrs. To better understand the delay, it is necessary to perform an analysis via models that are fed with data of the gas content and the metallicity evolution. Also, a deeper analysis of the dust and metallicity characteristics of these galaxies would provide some insight on the origin of this delay. {\sl SPICA} / SAFARI spectroscopy is probably the best instrument that could be used to carry out this work.

Figs.~\ref{FigAfuv} and \ref{FigSFRD} taken together at face value would suggest that the universe's dusty era (meaning dust attenuation higher than in the local universe) started at z = 3 - 4 simultaneously with the rise of a universe-wide star-formation event.

Figs.~\ref{FigSFRD} allowed us to follow the SFRD over most of the Hubble time in a consistent way. However, large uncertainties prevented us from closing the case. To go further in redshift, we need to detect and characterize galaxies at higher redshifts. This means better sensitivities and redshift measurements. {\sl SPICA} spectroscopy and {\sl SPICA} / Mid-IR instrument are best suited to extend the present work to much higher redshift.

 This work is further detailed in \citep[][]{Burgarella_2013}.


\bibliography{dburgarella}

\begin{thebibliography}{}
\expandafter\ifx\csname natexlab\endcsname\relax\def\natexlab#1{#1}\fi
\expandafter\ifx\csname url\endcsname\relax
  \def\url#1{\texttt{#1}}\fi
\expandafter\ifx\csname urlprefix\endcsname\relax\def\urlprefix{URL }\fi
\providecommand{\eprint}[2][]{\url{#2}}

\bibitem[{Barger et~al.(2012)}]{Barger_2012}
Barger, A.~J., et~al. 2012, ApJ, 761, 89

\bibitem[{Behroozi et~al.(2012)Behroozi, Wechsler, \& Conroy}]{Behroozi_2012}
Behroozi, P.~S., Wechsler, R.~H., \& Conroy, C. 2012, ApJ, 770, 57

\bibitem[{Bouwens et~al.(2009)}]{Bouwens_2009}
Bouwens, R.~J., et~al. 2009, ApJ, 705, 936

\bibitem[{Buat et~al.(2009)}]{Buat_2009}
Buat, V., et~al. 2009, A\&A, 507, 693

\bibitem[{Burgarella et~al.(2013)Burgarella, Buat, Gruppioni, Cucciati, Heinis
  et~al.}]{Burgarella_2013}
Burgarella, D., Buat, V., Gruppioni, C., Cucciati, O., Heinis, S., et~al. 2013,
  A\&A, 554, 70

\bibitem[{Burgarella et~al.(2005)}]{Burgarella_2005}
Burgarella, D., et~al. 2005, MNRAS, 360, 1413

\bibitem[{Chary \& Elbaz(2001)}]{Chary_2001}
Chary, R., \& Elbaz, D. 2001, A\& A, 556, 562

\bibitem[{Cucciati et~al.(2012)}]{Cucciati_2012}
Cucciati, O., et~al. 2012, A\& A, 539, 31

\bibitem[{Dwek \& Cherchneff(2011)}]{Dwek_2011}
Dwek, E., \& Cherchneff, I. 2011, ApJ, 727, 63

\bibitem[{Franceschini et~al.(2010)}]{Franceschini_2010}
Franceschini, A., et~al. 2010, A\& A, 517, 74

\bibitem[{Goto et~al.(2010)}]{Goto_2010}
Goto, T.~e., et~al. 2010, A\&A, 514, 6

\bibitem[{Gruppioni et~al.(2013)}]{Gruppioni_2013}
Gruppioni, C., et~al. 2013, MNRAS, 432, 23

\bibitem[{Hopkins \& Beacom(2006)}]{Hopkins_2006}
Hopkins, P.~F., \& Beacom, J. 2006, ApJ, 651, 142

\bibitem[{Le~Floc'h et~al.(2005)}]{LeFloch_2005}
Le~Floc'h, E., et~al. 2005, ApJ, 632, L169

\bibitem[{Lutz et~al.(2011)}]{Lutz_2011}
Lutz, D., et~al. 2011, A\&A, 532, L90

\bibitem[{Magnelli et~al.(2012)}]{Magnelli_2012}
Magnelli, B., et~al. 2012, A\&A, 539, 155

\bibitem[{Oliver et~al.(2012)}]{Oliver_2012}
Oliver, S.~J., et~al. 2012, MNRAS, 424, 1614

\bibitem[{Takeuchi et~al.(2005)Takeuchi, Buat, \& Burgarella}]{Takeuchi_2005}
Takeuchi, T.~T., Buat, V., \& Burgarella, D. 2005, A\&A, 440, L17

\bibitem[{Valiante et~al.(2009)Valiante, Schneider, Bianchi, \&
  Andersen}]{Valiante_2009}
Valiante, R., Schneider, R., Bianchi, S., \& Andersen, A.~J. 2009, MNRAS, 397,
  1661

\end{thebibliography}

\end{document}